# Novel temperature operated electrooptical materials based on $K_2ZnCl_4$: PVA polymer nanocomposites

A.M. El-Naggar [a,*], A.A. Albassam [a], J. Jedryka [b], A. Wojciechowski [b]

[a] *Research Chair of Exploitation of Renewable Energy Applications in Saudi Arabia, Physics & Astronomy Department, College of Science, King Saud University, P.O. Box 2455, 11451 Riyadh, Saudi Arabia*
[b] *Institute of Optoelectronics and Measuring Systems, Faculty of Electrical Engineering, Czestochowa University of Technology, Armii Krajowej 17, Czestochowa, Poland*

**Abstract.** We have discovered significant temperature anomalies of the linear electrooptical (EOE) coefficients for the $K_2ZnCl_4$ nanocrystallites embedded into the polymer matrices. The EOE efficiency was operated by external dc-electric field and UV polarized laser light. The maximal enhancement of EOE output was achieved within the temperature range 375 K–420 K. This temperature range corresponds to existence of incommensurate structural modulation. The maximally achieved field induced EOE efficiency was equal to about 1.4 pm/V for off-diagonal EOE tensor component. The size and NC content dependences have been explored for the size range (20 nm–1500 nm) and contents varying within the 25%–35%. For the sizes about 1200 nm, the EOE is commensurable with the EOE of the bulk crystals and is equal to about 0.6 pm/V.

## 1. Introduction

Recently one can observe an increasing effort to search and design of new materials possessing linear EOE response. This is crucial for design of laser modulators, deflectors and triggers. The EOE application is determined by second order optical susceptibility tensors described by third rank polar tensors [1–3]. It is important to point out that many optical effects of second and third order are not only observed for the single crystals, however they exist also in fine-grained ceramics [4,5], meta materials [4] and other disordered media. These materials may be used for nonlinear optical wave mixing, optical triggering, laser modulators, deflectors, memory cells etc.

Among these compounds particular interest present materials possessing ferroic phase [5] including incommensurate structural modulation. Growth of pure and doped $K_2ZnCl_4$(KZC) single crystals was carried out by traditional Czochralski technique[6]. The KZC single crystals were grown by Czochralski technique. Their structure was monitored by X-ray diffraction method. The grown KZC single crystals exhibited an inclusion of the high-temperature *Pmcn* phase with lattice parameters $a$ = 7.263(2) Å, $b$ = 12.562(2) Å and $c$ = 8.960(4) Å into the *P21cn* ambient temperature stable phase. The results of the birefringence studies confirmed also an appearance of incommensurate phase structural modulation. The existence of incommensurate structure is responsible for the occurrence of additional nonlinear optical features [7] due to addition of long range ordering modulation. This one opens a rare possibility for varying their optoelectronic features using external laser beams [8]. Very important here is an occurrence of superstructure of superimposing several simple lattices. The coexistence of lattice structures with different lattice parameters may lead to an enhanced nonlinear optical response. The more appropriate is their use in a form of nanocrystallites embedded into the photopolymer matrices [9].

## 2. Experimental methods

The titled crystals of KZC have been grown by traditional Czochralski technique in an argon atmosphere, using an experimental setup allowing direct visual observation to the whole grown zone. Slowly cooled crystals exhibited excellent cleavage mechanical properties. The mechanically and acoustically milled samples of the nanocrystallites have been separated to several groups with respect to their sizes. Using TEM methods, the nanocrystallites with average sizes 5 nm, 12 nm, 25 nm, 50 nm 80 nm, 120 nm and 250 nm were chosen. After they have been embedded into the polyvinyl alcohol (PVA) liquid photopolymers with photo-initiators of different content. This phase has been photosolidified using external UV laser light at 371 nm wavelength

* Corresponding author.
*E-mail address:* elnaggar@ksu.edu.sa (A.M. El-Naggar).

[10] during 3–5 min. The PVA matrices have strong benefits with respect to the other ones due to their excellent piezooptical features. The UV-solidification was performed during simultaneously applied external dc-electric field with electric strengthup to 8 kV/cm and UV polarized light at λ = 371 nm with power density equal to about 45 W/$cm^2$. The control of time which is necessary for photosolidification was done using the interferometric patterns in the cross polarized light. The birefringence was measured by traditional Senarmont method using 1150 nm cw He-Ne laser of power about 25 mW. The beam diameter was varied within the 3–6 mm. The accuracy of birefringence determination was equal to about $10^{-5}$. Because the nanocrystallites in the polymer matrices were treated simultaneously by UV polarized light and dc-electric field aligned, we consider two possible geometries. The first one: parallel geometry, where the dc-electric field and the probing beam polarization were parallel (diagonal), the second one:perpendicular directions (off-diagonal). Our experiments have shown that maximal changes of birefringence under external dc-electric and optical fields were atleast one order higher for the diagonal geometry. Typical dependences of the principal EOE coefficients versus the applied dc-electric field for the nanocomposites for two principal light polarizations at optimal UV-polarized power density 45 W/$cm^2$ are presented in the Fig. 1.

## 3. Electrooptical results

Dependence of the EOE at λ = 1150 nm for the studied composites versus temperature is shown in the Fig. 2. The experimental data are shown for optimal magnitudes of applied dc-electric fields and UV-induced laser power densities. The main obtained results, consist in an occurrence of a huge EOE enhancement within the temperature range between375 K and 420 K. These temperatures correlate well with the temperature range of incommensurate phase existence. The observed EOE increases up to 3.5 times. This one may be used for application as EOE light modulators. In this case one can operate by EOE with simultaneous variation of content and temperature contrary to the traditional polymer EOE materials [11,12].

Following the presented data one can see a crucial dependence on temperature. The EOE values have been observed for the whole range of the existed incommensurate phase and their maxima correlates well with the maximal structural modulation. They are significantly higher than for the traditional ferroelectric materials. The enhancement of the effect due to incommensurate phase may be caused by occurrence of long range ordering which give additional non-centrosymmetric space charge density necessary for enhancement of the linear electrooptics. So we achieved an enhanced linear EOE due to occurrence of incommensurate superstructure. Its magnitude is high with respect to the pure ferroelectric transitions. Moreover, during the UV-solidification of the titled NC, there occur some interferometric patterns observed in cross polarizes light. They are caused by NC-polymer interfaces with thickness of about 2–3 nm as shown in the Fig. 3 for different geometries of the light and applied fields. These patterns confirm a prevailing role of the non-solidified thin films defining the interferometric monitoring.

The obtained results unambiguously demonstrate a huge potential of the NC possessing incommensurate phases for application as polymer light modulators. The effects disappeared after switching off the electrical or optical UV-illumination within the 120…160 ms (see Fig. 4). These materials show huge stability (with reproducibility of the EOE up to 0.6%) in time. After the treatment with dc-field and UV light, it was achieved a reproducibility up to 500 cycles. These effects are slightly dependent on the light polarization and UV solidified beam profiles. It is maximal for such time of UV solidification during which we have some thin nanosheets of non-solidified range. The degree of scattering light did not exceed 2%. The performed studies indicate that principal origin of the EOE effect is caused by occurrence of incommensurate modulation. The non-homogeneity of the polymer surfaces did not exceed 4.5%. The interfaces thickness defines an existence of the non-solidified sheets which allow to operate by effective EOE. The latter is also sensitive to partial aggregation of the KZC NC.

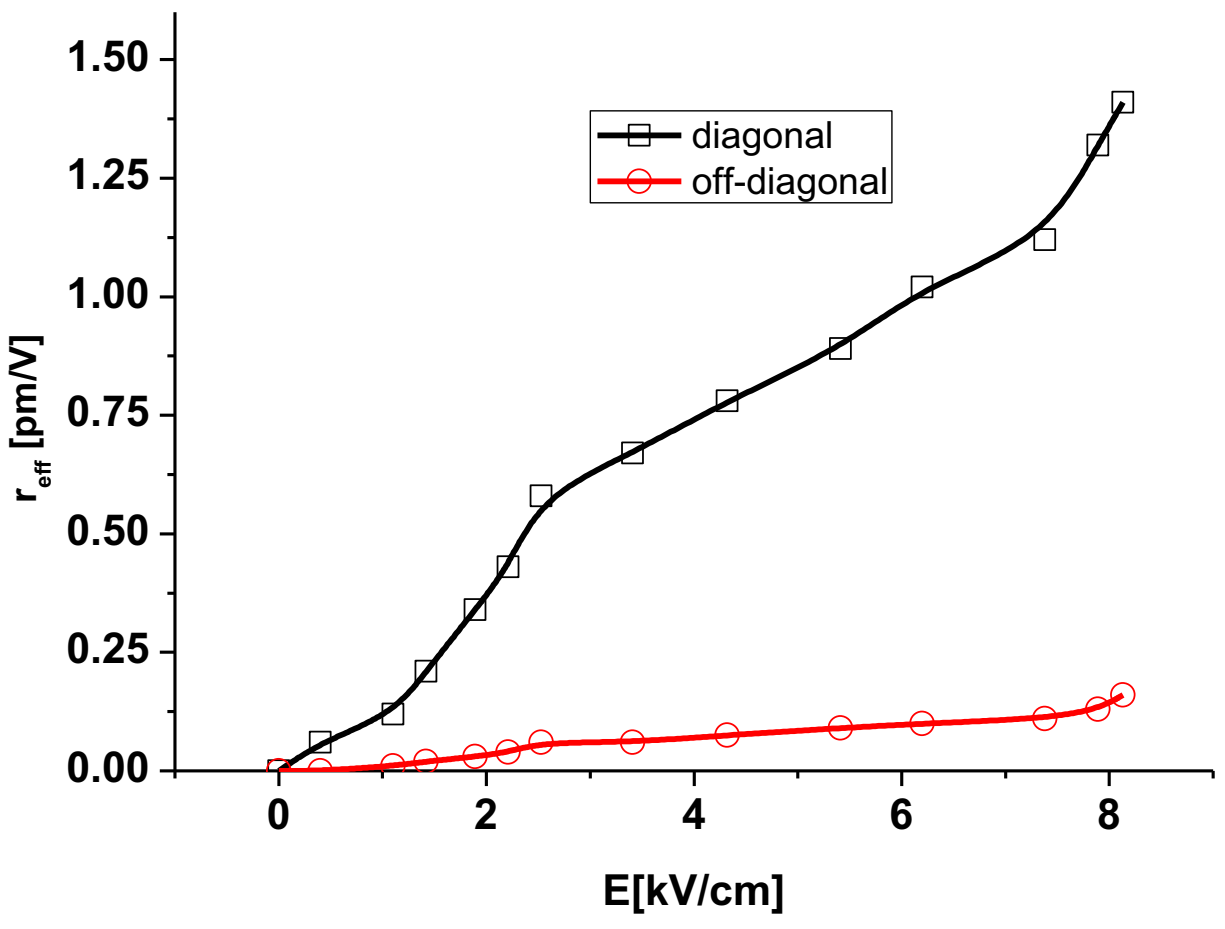

**Fig. 1.** Dependence of the EOE coefficients versus the applied dc-electric field for two principal directions of polarization of photoinduced UV laser beam. The optimal NC sizes corresponded to 25 nm and NC content of 33%.

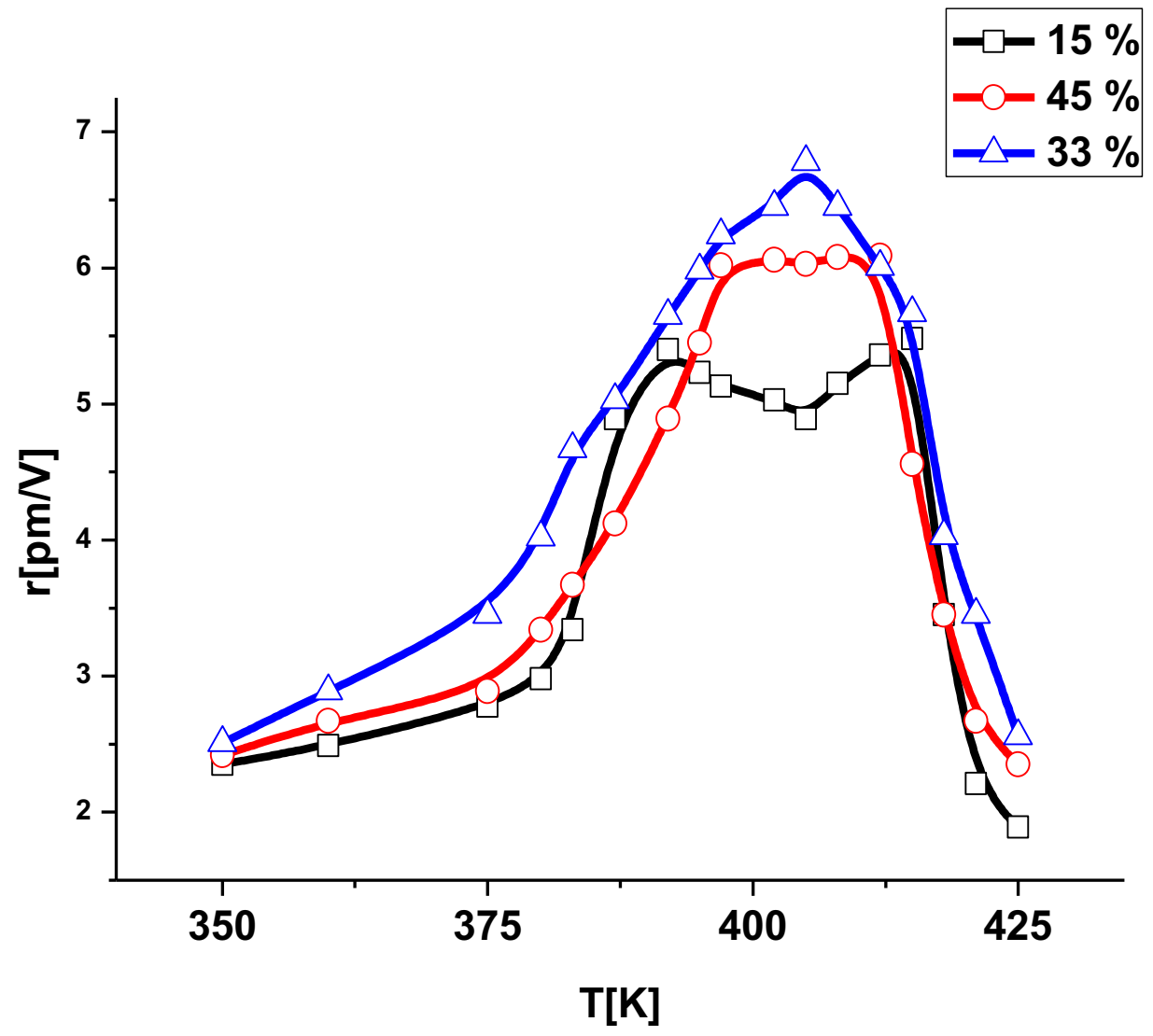

**Fig. 2.** Temperature dependence of the linear EOE at 1150 nm for the composites with different nanocrystallite content and sizes of about 20 nm at optimal external dc-electric field strengths and UV polarized power densities.

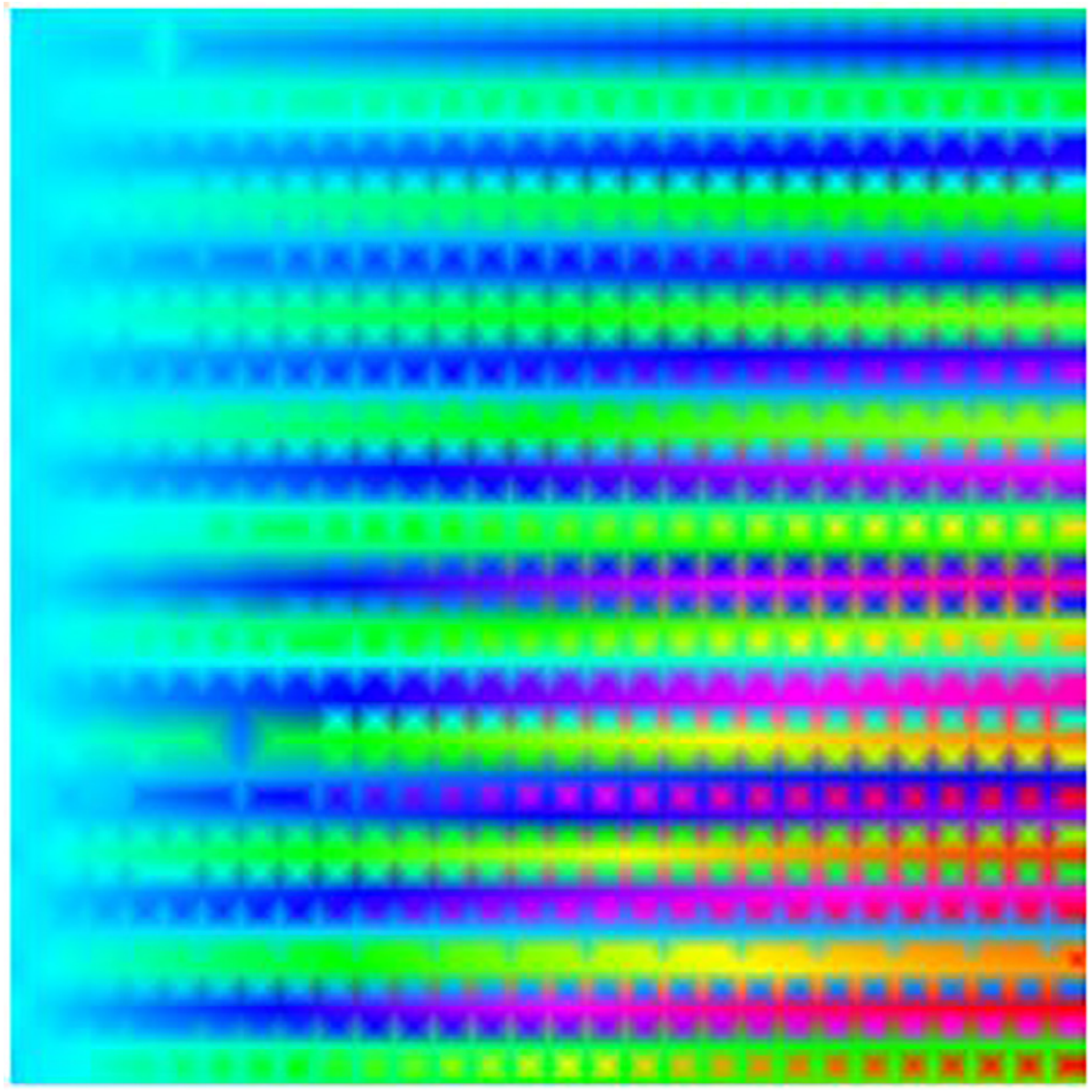

**Fig. 3.** Interferometer images of the studied compounds in the cross polarized light during simultaneous optical and dc-electric field treatment.

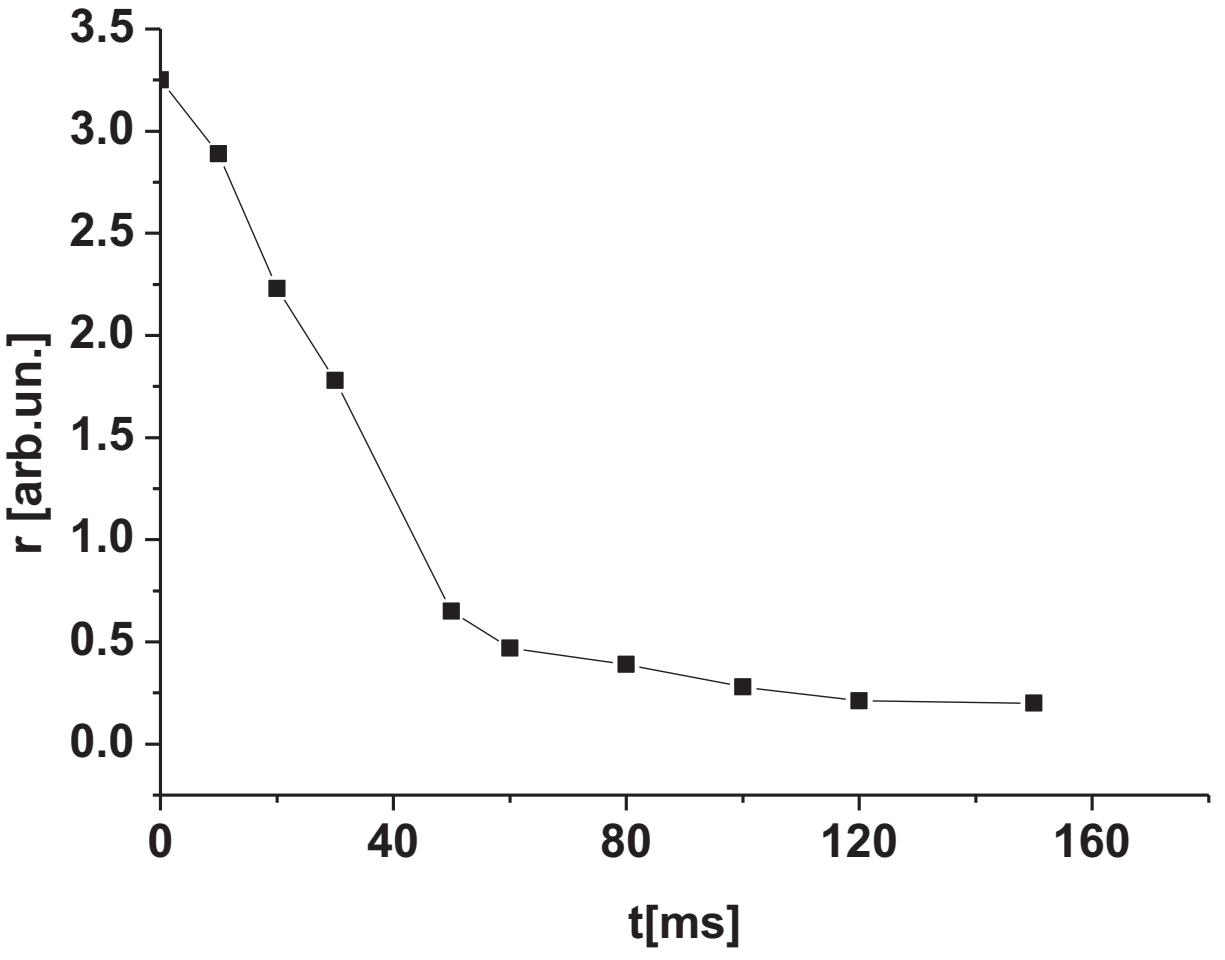

**Fig. 4.** Time decay of the EOE coefficients after switching off the external fields.

## 4. Conclusions

We have discovered a huge enhancement of EOE efficiency for the new synthesized $K_2ZnC_4$ NC embedded into the PVA photopolymer matrices. The maximal EOE corresponds to the content of nanocrystallites which is about 33% at average sizes equal to about 25 nm. Maximal enhancement of the linear EOE has been observed within the temperature range of 375 K–420 K. This effect exists only during simultaneous dc-electric and UV laser treatments and disappears after switching off the external fields during the range of 120–160 ms. It is significantly higher than that of the traditional ferroelectric materials.

## Acknowledgements

The authors are grateful to the Deanship of Scientific Research, King Saud University for funding through Vice Deanship of Scientific Research Chairs.

For J.J. and A.W. this work is a part of a project European Union's Horizon 2020 research and innovation program under the Marie Skłodowska-Curie grant agreement No 778156.